\documentclass[11pt]{article}

\usepackage[numbers]{natbib}
\usepackage{amsmath, amsthm, amssymb, amsfonts, multirow,hyperref}
\usepackage{bbm}
\usepackage{xargs,setspace}
\usepackage{amsfonts}

\usepackage[small,bf]{caption}

\usepackage{fullpage}
\usepackage{graphics,url,verbatim,float}
\usepackage{graphicx}
\usepackage{algorithm,algorithmic}
\usepackage[parfill]{parskip}
\usepackage{subcaption}

\usepackage{xargs}
\usepackage{amsfonts}

\usepackage{color}

\newcommand{\PrimalFull}{\texttt{Primal-Full}}
\newcommand{\DualFull}{\texttt{Dual-Full}}

\newcommand{\M}{\mathbf}

\newcommand{\sbt}{\mathrm{s.t. }}

\newcommand{\DSL}{\mbox{$\ell_{1}$-DS}}
\newcommand{\DSLp}{\mbox{$\ell_{1}$-DS}}

\newcommand{\RM}[1]{\textcolor{blue}{[RM] #1}}

\usepackage{algorithm}
\usepackage{lscape}

\DeclareMathOperator*{\argmin}{arg\,min}

\DeclareMathOperator*{\mini}{minimize}
\DeclareMathOperator*{\maxi}{maximize}

\usepackage{array,graphicx}
\usepackage{booktabs}
\usepackage{pifont}

\begin{document}

\title{Computing Estimators of Dantzig Selector type via Column and Constraint Generation}

\author{Rahul Mazumder\thanks{MIT Sloan School of Management, Operations Research Center and Center for Statistics, MIT.}~~~~Stephen Wright\thanks{Department of Computer Sciences,
University of Wisconsin-Madison.}~~~~Andrew Zheng\thanks{Operations Research Center, MIT.}}

\maketitle

\begin{abstract}
We consider a class of linear-programming based estimators in reconstructing a sparse signal from linear measurements. Specific formulations of the reconstruction problem considered here include Dantzig selector, basis pursuit (for the case in which the measurements contain no errors), and the fused Dantzig selector (for the case in which the underlying signal is piecewise constant). In spite of being estimators central to sparse signal processing and machine learning, solving these linear programming problems for large scale instances remains a challenging task, thereby limiting their usage in practice. We show that classic constraint- and column-generation techniques from large scale linear programming, when used in conjunction with a commercial implementation of the simplex method, and initialized with the solution from a closely-related Lasso formulation, yields solutions with high efficiency in many settings. 

\end{abstract}

%

\section{Introduction}
\label{section:introduction}
We consider the prototypical problem of sparse signal recovery from linear measurements~\cite{candes2007,tibshirani1996regression,chen1994}: given a model matrix $X \in \mathbb{R}^{n \times p}$ with $n$ samples and $p$ features, response 
$y \in \mathbb{R}^{n}$ generated via the model 
$y = X\beta^0 + \epsilon$, where, $\beta^0\in \mathbb{R}^{p}$ is sparse (that is,  has few nonzero entries) and the errors are i.i.d. Gaussian with mean zero and variance $\sigma^2$ (i.e., $\epsilon \sim N(0,\sigma^2 I)$). We consider the case in which the number of variables is much larger than the number of samples  ($p \gg n$)  and our task is to estimate $\beta^0$ from $(y, X)$, exploiting the knowledge that $\beta^0$ is sparse. 

We assume throughout that the columns of $X$ have been standardized to have mean zero and unit $\ell_{2}$-norm. The $\ell_1$-norm $\| \beta \|_1$ is often used as a convex surrogate to the cardinality of $\beta$, which is a count of the number of nonzero elements in $\beta$. The celebrated Dantzig Selector~\cite{candes2007} approximates $\beta^0$ by minimizing $\|\beta\|_1$ subject to a constraint on the maximal absolute correlation between the features and the vector of residuals (given by $r:= y - X\beta$). 
The optimization problem of this recovery problem is as follows: 
\begin{equation}\label{eq:dantzig}
\text{($\ell_1$-DS)} ~~~~~ \mini_{\beta} ~~~ \| \beta \|_{1} ~~~ \sbt ~~~ \| X^T(y - X\beta ) \|_{\infty} \leq \lambda, 
\end{equation} 
where $\lambda>0$ controls the data-fidelity term. Ideally, the value of $\lambda$ should be such that the unknown signal vector $\beta^0$ is feasible, that is, $\| X^T \epsilon \|_{\infty} \leq \lambda$ holds (with high probability, say). The constraint in \eqref{eq:dantzig} can be interpreted as the $\ell_{\infty}$-norm of the gradient of the least squares loss $\frac12\| y - X\beta \|_{2}^2$. An appealing property of the Dantzig Selector estimator is that it is invariant under orthogonal transformations of $(y,X)$. 

Problem \eqref{eq:dantzig} can be reformulated as a linear program (LP), and thus solved via standard LP algorithms and software (for example, commercial solvers like Gurobi and Cplex) for instances of moderate size. As pointed out in \cite{becker2011}, efficient algorithms for $\DSL$ are scarce:
 \begin{quote}
    ``...Typical modern solvers rely on interior-point methods which are somewhat
problematic for large scale problems, since they do not scale well with size." 
 \end{quote}
Although important progress has been made on algorithms for $\DSL$ in subsequent years (see, for example, \cite{becker2011,lu2012,wang2012,Pang:fastclime-package,JMLR:v16:li15a}), large-scale instances of \eqref{eq:dantzig} (with $p$ of a million or more) still cannot be solved. The main goal of our work is to improve our current toolkit for solving $\DSL$ and related problems, bringing to bear some underutilized classical tools from large scale linear programming.  

The Dantzig Selector is closely related to the Lasso~\cite{tibshirani1996regression}, which combines a least squares data-fidelity term with an $\ell_1$-norm penalty on $\beta$. While the Lasso and Dantzig Selectors yield different solutions~\cite{efron2007discussion},
for suitably chosen regularization parameters, they both lead to estimators with similar statistical properties in terms of estimation error, under suitable assumptions on $X$, $\beta^0$, and $\sigma$ (see \cite{bickel2009simultaneous}). A version of the Lasso that has the same objective as \eqref{eq:dantzig} is
\begin{equation}\label{lasso-DS-form}
\mini_{\beta}~~\| \beta \|_{1} ~~ \sbt ~~ \| y - X\beta \|_{2} \leq \delta,
\end{equation}
where $\delta \geq 0$ is a parameter that places a budget on the data fidelity, defined here as the $\ell_2$-norm of the residuals. 
The explicit constraint on data fidelity makes this formulation appealing, but it poses computational challenges~\cite{becker2011} because of the difficulty of projecting onto the constraint. The following alternative, unconstrained version has become the most popular formulation of Lasso: 
\begin{equation}
\label{eq:lasso}
\text{(Lasso)}~~~~\mini_{\beta}~~\frac12 \| y - X\beta \|_{2}^2 + \lambda \| \beta \|_{1}
\end{equation}
where $\lambda \geq 0$ is a regularization parameter that controls the $\ell_1$-norm of $\beta$. (This is the formulation that we refer to as ``Lasso'' in the remainder of the paper.)
There are several highly efficient algorithms for \eqref{eq:lasso} (see, for example, \cite{beck2009,friedman2010,wright2009sparse}), making it an extremely effective tool in the context of sparse learning based on $\ell_1$-minimization. 

\subsection{Algorithms for Lasso and Dantzig Selector: Fixing the Gap in Performance.} 

Common algorithms for solving~\eqref{eq:lasso} are based on proximal gradient methods~\cite{beck2009,wright2009sparse}, coordinate descent~\cite{friedman2010}, or homotopy methods~\cite{efron2004least}. Several efficient implementations of these methods are available publicly. There are key differences between the Lasso and $\DSL$ in terms of computational properties and associated solvers: $\DSL$ is essentially an LP whereas Lasso is a convex quadratic program (QP). Although LPs are generally thought to be easier to solve than QPs of a similar size, the Lasso QP can be solved with remarkable efficiency, at least when $\beta$ is quite sparse and the matrix $X$ has felicitous properties.


While first order optimization algorithms~\cite{becker2011,boyd2011} have also led to good algorithms to solve $\DSLp$, they are still much slower than Lasso. 
To illustrate, to compute a path of 100 solutions for a problem with $n=200$, $p=12,000$, \texttt{glmnet}~\cite{friedman2010} takes $0.24$ seconds with minimal memory requirement on a modest desktop computer. On the other hand, for the same dataset (and machine), solving $\DSL$ for a path of 100 $\lambda$ values by the parametric simplex method of~\cite{Pang:fastclime-package} takes several minutes and requires at least $10$GB of  memory. The software package \texttt{flare}~\cite{JMLR:v16:li15a}, based on the Alternating Direction Method of Multipliers (ADMM)~\cite{boyd2011} has prohibitive memory requirements and would not run on a modest desktop machine. 
The differences between solvers grow with problem size. As a consequence of the difficulties of solving the LP formulation, $\DSL$ remains somewhat under-utilized, in spite of its 
excellent statistical properties.
This paper seeks to address the striking difference in computational performance between the Lasso and the Dantzig Selector by proposing efficient methods for the latter.
We make use of classical techniques from optimization: column generation and constraint generation. These techniques were first proposed as early as 1958~\cite{ford1958suggested,dantzig1960decomposition} in the context of solving large scale LPs but, to our knowledge, have not been applied to $\DSL$ or its relatives discussed below. 

Our approach exploits the sparsity that is typically present in the solution of $\DSL$: at optimality, an optimal $\beta$ will have few nonzeros. If we can identify the nonzero components efficiently, we may avoid having to solve a full LP formulation that includes all $p$ components of $\beta$. Column generation starts by selecting a subset of components in $\beta$ and solving a reduced version of \eqref{eq:dantzig} that includes only these components (that is, it fixes the components of $\beta$ that are {\em not} selected to zero). If the optimality conditions for the full problem are not satisfied by the solution of the reduced LP, more components of $\beta$ are added to the formulation in a controlled way, and a new reduced LP is solved, using the previous solution as a warm start. The process is repeated until optimality for the full LP is obtained. Whenever new components are added to the reduced LP, we add new columns to the constraint matrix, hence the name {\em column generation}.

We make use too of another key property of $\DSL$: redundancy of the constraints in \eqref{eq:dantzig}. Typically, the number of components of $X^T(y-X\beta)$ that are at their bounds of $-\lambda$ and $\lambda$ at the solution is small, of the same order as the number of nonzero components of $\beta$ at the solution.  This observation suggests a procedure in which we solve a reduced LP with just a subset of constraints enforced. We then check the constraints that were not included in this formulation to see if they are violated by the solution of the reduced LP. If so, we add (some of) these violated constraints to the LP formulation, and solve the modified reduced LP. This process is repeated until a solution of the original problem is obtained. This procedure is known as {\em constraint generation}.


While column generation and constraint generation are commonly used as separate entities to solve large scale LPs, it makes sense to use them jointly in solving $\DSL$, and a combination of the two strategies can be implemented with little complication. The procedures can benefit from good initial guesses  of the nonzero components of $\beta$ and the active constraint set for \eqref{eq:dantzig}. We use efficient Lasso solvers to obtain these initial guesses.  



\subsection{Other Examples}
Several other examples of sparse linear models are also amenable to column and constraint generation techniques. These include basis pursuit denoising~\cite{chen1994} and a Dantzig selector version of one-dimensional total variation denoising (also known as fused Lasso)~\cite{mammen1997locally,tibshirani2005}. Each of these problems can be formulated as a linear program task and, like $\DSL$, they are computationally challenging.

\paragraph{Basis Pursuit.} 
The noiseless version of $\ell_1$ sparse approximation, popularly known as Basis Pursuit~\cite{chen1994}, is given by the following optimization problem:
\begin{equation}\label{eq:basis_pursuit}
\mini_{\beta} ~~ \| \beta \|_{1} ~~~~ \sbt ~~~ y = X\beta,
\end{equation}
which can be formulated as an LP. 
This problem can be interpreted as a limiting version of \eqref{eq:lasso} as $\lambda \rightarrow 0+$.
It may be tempting to solve \eqref{eq:lasso} for a small value of $\lambda \approx 0$ (possibly with warm-start continuation) to obtain a solution to~\eqref{eq:basis_pursuit}. 
However, this approach is often inefficient in practice, because obtaining accurate solutions to the Lasso becomes increasingly expensive as $\lambda \downarrow 0$.
It has been pointed out in \cite{donoho2009message} that solving \eqref{eq:basis_pursuit} to high accuracy using existing convex optimization solvers or specialized iterative algorithms is a daunting task, for large instances. Our own experiments show that current solvers based on ADMM fail to solve \eqref{eq:basis_pursuit} for $p \geq 10^4$ (with  $n<1000$), while our proposed approach, described below,
solves problems with $p \approx 4 \times 10^5$  within 3-4 minutes. 
Our approach relies on column generation, exploiting the familiar observation that the solution $\beta$ of \eqref{eq:basis_pursuit} is sparse.

\paragraph{Fused Dantzig Selector.} 
The Fused Lasso~\cite{tibshirani2005} or the total variation penalty~\cite{rudin1992nonlinear} is a commonly used  $\ell_1$-based penalty that encourages the solution to be (approximately) piecewise constant. 
The unconstrained  formulation of this problem is
\begin{equation}\label{eq:fused_lasso}
\mini_{\beta}~~~\frac12 \| y - X \beta \|_{2}^2 + \lambda \| D^{(0)} \beta \|_{1},
\end{equation}
where $\lambda\geq 0$ is a regularization parameter and $D^{(0)} \in \mathbb{R}^{(p-1) \times p}$ is the first order difference operator 
matrix, defined by
\begin{equation} \label{def.D0}
    D^{(0)}\beta = (\beta_{2}  - \beta_{1}, \beta_{3} - \beta_{2}, \ldots, \beta_{p} - \beta_{p-1})^T,
\end{equation}
 which represents  differences between successive components of $\beta$. 
As we show in Section~\ref{section:fused_dantzig}, \eqref{eq:fused_lasso} can be expressed as a  Lasso problem of the standard form \eqref{eq:lasso}, with a modified model matrix $\tilde{X} \in \mathbb{R}^{n \times p-1}$ and response $\tilde{y} \in \mathbb{R}^n$~\footnote{We define $(\tilde{y},\tilde{X})$ as follows. Define 
$D=[e_{1}^T; D^{(0)}] \in \mathbb{R}^{p\times p}$, where $e_1=(1,0,0,\dotsc,0)^T$, and define $H=D^{-1}$.  For any $A \subset \{1,2,\dotsc,p\}$,  let $H_{A}$ be the submatrix of $H$ containing the columns indexed by $A$, and let $P_{A}$ denote the projection operator onto the column space of $H_{A}$. We set $\tilde{y} := (I - P_A) y$ and $\tilde{X} := (I - P_A) XH_B$, with $A=\{1\}$ and $B=\{2, \ldots, p\}$.}. This suggests a natural Dantzig Selector analog of the fused Lasso problem:
\begin{equation}
\label{eq:dantzig_tf_proj}
\mini_{\alpha  \in \mathbb{R}^{p-1}}~~\|\alpha \|_1~~\sbt~~\| \,  \tilde{X}^T(\tilde{y} - \tilde{X} \alpha) \|_\infty \leq \lambda,
\end{equation} 
a formulation that is  amenable to the column and constraint generation methods developed in this paper. In the special case of $X=\mathbb{I}$, the problem has additional structure that we can exploit to solve  instances with $p\approx 5 \times 10^5$, well beyond the capabilities of alternative methods.

\subsection{Related work and Contributions}


The Dantzig Selector formulations presented here  --- \eqref{eq:dantzig}, \eqref{eq:basis_pursuit}, and \eqref{eq:dantzig_tf_proj} ---
can all be expressed as LPs and solved with interior point methods or simplex-based methods, 
as implemented in commercial solvers \footnote{Commercial solvers such as Gurobi, Cplex, Mosek are free for academic use.} (Gurobi, Cplex, Mosek, XPress, etc) or open-source codes (GLKP, lpsolve, among others). 
Specialized implementations for $\DSL$ and Basis Pursuit have been investigated for several years. 
An interior point method was used  in \cite{candes2007}, a first-order method for a regularized version of $\DSLp$ was described in \cite{becker2011}\footnote{The authors add a small ridge penalty to the objective and optimize the dual via gradient methods}, and methods based on ADMM were discussed in \cite{lu2012,wang2012,boyd2011}.
Using a homotopy continuation approach, \cite{james2009} extend the framework of LARS~\cite{efron2004least} to find the solution path to  $\DSLp$, which is piecewise linear in $\lambda$.
Homotopy continuation methods applicable to $\DSL$ have also been proposed by \cite{asif2009,brauer2018primal,pang2017}, but these works do not appear to use column and 
constraint generation methods, which are the main focus of our work. 
For the $\DSL$ problem, the algorithms of~\cite{asif2009,brauer2018primal,pang2017} compute the full $p\times p$ matrix $X^TX$ at the outset. This operation is memory-intensive, so these  approaches can handle values of $p$ only up to a few thousands on a modest desktop computer.
Our methods solve the problems \eqref{eq:dantzig}, \eqref{eq:basis_pursuit}, and \eqref{eq:dantzig_tf_proj} at a single value of the regularization parameter, but they can be extended to solve these problems on a grid of regularization parameters via a warm start continuation strategy. We show that 
the classical tools of column and constraint generation can be effective in solving large-scale instances of these problems.  Our work is related to the proposal of~\cite{dedieu2019solving} who explored column and constraint generation to solve 
regularized linear SVM problems (with a hinge loss) that can be expressed as LPs. (Regularizers considered in \cite{dedieu2019solving} include the $\ell_1$-norm, group $\ell_1$-norm, and the Slope penalty.) Our  Dantzig Selector
problems have structural properties different from the SVM problems. They also have the unique advantage that they can be initialized using Lasso. This fact plays an important role in the practical computational efficiency of our approaches.



Our methods are based on the simplex algorithm, which is better at making use of the available warm-start information than interior-point methods. A memory-friendly version of Gurobi's simplex solver, applied to an LP formulation of \eqref{eq:dantzig} that avoids formation of $X^TX$ by using auxiliary variables, works well for $\DSL$ with $n$ in the  hundreds and $p$ in the thousands. In fact, this approach can be  faster than some specialized algorithms~\cite{pang2017,becker2011,lu2012}.
We show simplex performance can be improved substantially by using column and constraint generation when $p \gg n$, for problems in which the underlying solution is sufficiently sparse.
We refer to our framework as 
Dantzig-Linear Programming (DantzigLP for short). 
Because we use a simplex engine as the underlying solution, a primal-dual solution is available at optimality. If we decide to terminate the algorithm early due to computational budget constraints, our framework delivers a certificate of suboptimality. 
DantzigLP can solve instances of the 
$\DSL$ problem with $n\approx 10^3$ and $p \approx 10^6$;
Basis Pursuit with $n\approx 10^3$ and $p \approx 10^5$; and  Fused Lasso with $n=p \approx 10^6$; all within a few minutes and with reasonable memory requirements.
To our knowledge, problems of this size are beyond the capabilities of current solvers.
A Julia implementations of our DantzigLP framework can be found at
\url{https://github.com/atzheng/DantzigLP}.

\paragraph{Notation.}

We denote $[n] := \{1, 2, \ldots n\}$. The identity matrix is denoted by $\mathbb{I}$  (with dimension understood from the context).
When operating on vector-valued operands $u, v \in \mathbb{R}^n$, the inequality $u \leq v$ denotes elementwise comparison.
For any matrix $X \in \mathbb{R}^{n \times p}$ and index sets $I \subset [n]$ and $J \subset [p]$, we denote by $X_{I, J}$ the $|I| \times |J|$ submatrix of $X$ that consists of the rows of $X$ indexed by $I$ and the columns of $X$ indexed by $J$.  The notation 
$X_{*, J}$ denotes a submatrix consisting of all rows of $X$ but only the columns indexed by $J$ (A similar convention applies to
$X_{I, *}$). For a vector $v \in \mathbb{R}^n$ and a set $S \subset [n]$, $v_{S}$ denotes the subvector of $v$ restricted to the indices in 
$S$. The notation $\M{1}$ denotes the vector $(1,1,\dotsc,1)^T$, whose length is defined by the context.

\section{Methodology}

\subsection{Column and Constraint Generation for Large-Scale LP}


{\em Column generation}~\cite{ford1958suggested,dantzig1960decomposition,bertsimas1997} is a classical tool to solve large scale LPs with a large number of variables and a relatively small number of constraints, when we anticipate an optimal solution with few nonzero coefficients. 
The basic idea is simple: we solve a small LP involving just
a subset of columns, and incrementally add columns into the model, re-solving the LP after each addition, until optimality conditions for the original problem are satisfied.
{\em Constraint generation} is used when the number of constraints is large relative to the number of variables, when we expect a relatively small subset of the constraints to be active at optimality.

For the sake of completeness, we provide an overview of these techniques in this section, referring the reader to \cite{bertsimas1997} for a more detailed treatment.

Given problem data  $A \in \mathbb{R}^{m \times n},$ $b \in \mathbb{R}^m$ and $c \in \mathbb{R}^n$ with decision variable $x \in \mathbb{R}^n$, we 
consider the LP~\eqref{eq:mfull}, whose dual~\eqref{eq:mfull-dual} has decision variable $v \in \mathbb{R}^m$.  We assume that $A$ has full rank. 

\begin{minipage}[t]{.48\linewidth}
\begin{equation}
    \label{eq:mfull}
    \tag{\PrimalFull}
    \begin{aligned}
        & \underset{x}{\text{minimize}} 
        & & c^Tx \\
        & \sbt  & & Ax \geq b \\
        & & & x \geq 0 
    \end{aligned}
\end{equation}
\end{minipage}
\begin{minipage}[t]{.02\linewidth}
~~~~~~~~~~
\end{minipage}
\begin{minipage}[t]{.48\linewidth}
\begin{equation}
    \label{eq:mfull-dual}
    \tag{\DualFull}
    \begin{aligned}
        & \underset{v}{\text{maximize}} 
        & & b^Tv \\
        & \sbt  & & A^T v \leq c \\
        & & & v \geq 0.
    \end{aligned}
\end{equation}\end{minipage}

We will assume that~\eqref{eq:mfull} has a finite optimal solution. By LP duality theory, the dual also has a solution with the same optimal objective value.

The solutions of \eqref{eq:mfull} and \eqref{eq:mfull-dual} can be derived from the solutions to reduced problems of the following form, for some index sets $I \subset [m]$ and $J \subset [n]$: \newline 
\begin{minipage}[t]{.45\linewidth}
\begin{equation}
    \label{eq:mred}
    \tag{\texttt{Primal}$(I, J)$}
    \begin{aligned}   
        & \mini_{x_J}
        & & c_{J}^Tx_{J} \\
        & \sbt  & & (A_{I, J}) x_{J} \geq b_I \\
        & & & x_J \geq 0, 
    \end{aligned}
\end{equation}
\end{minipage}~~
\begin{minipage}[t]{.02\linewidth}
~~~~~~~
\end{minipage}
\begin{minipage}[t]{.45\linewidth}
\begin{equation}
    \label{eq:mred-dual}
    \tag{\texttt{Dual}$(I, J)$}
    \begin{aligned}
        & \maxi_{v_I}~~
        & & b_I^T v_I \\
        & \sbt  & & A_{I, J}^T v_I \leq c_J \\
        & & & v_I \geq 0.
    \end{aligned}
\end{equation}\end{minipage}

The subsets $I$ and $J$ are not known in advance; the simplex method can be viewed as a search for these subsets in which typically one element is changed at each iteration. Sufficient conditions for the solutions $x_J$ of \eqref{eq:mred} and $v_I$ of \eqref{eq:mred-dual} to be extendible to solutions of \eqref{eq:mfull} and \eqref{eq:mfull-dual} are that:
\begin{equation} \label{eq:su8}
A_{i,J}x_J \ge b_i, \;\; \mbox{for all $i \in [m] \setminus I$}; \quad
A_{I,j}^T v_I \le c_j, \;\; \mbox{for all $j \in [n] \setminus J$}.
\end{equation}
If these conditions are satisfied, we obtain solutions $x^*$ and $v^*$ of \eqref{eq:mfull} and \eqref{eq:mfull-dual}, respectively, by setting $x^*_J=x_J$ and $x^*_{J^c} = 0$, and $v^*_I=v_I$ and $v^*_{I^c} = 0$.

Column and constraint generation are techniques for systematically expanding the sets $I$ and $J$ until the optimality conditions \eqref{eq:su8} are satisfied by the solutions of \eqref{eq:mred} and \eqref{eq:mred-dual}. At each iteration, we solve a reduced problem of this form, then seek indices $i$ and $j$ for which the conditions \eqref{eq:su8} are violated. Some of these indices are added to the sets $I$ and $J$, and the new (slightly larger) versions of \eqref{eq:mred} and \eqref{eq:mred-dual} are solved using a variant of the simplex method, typically warm-started from the previous reduced problem. An  outline of the approach is shown in Algorithm~\ref{alg:ccgen}.

\begin{algorithm} 
\caption{Constraint and Column Generation to solve \eqref{eq:mfull} 
and \eqref{eq:mfull-dual}}\label{alg:ccgen}
\begin{algorithmic}
\STATE Initialize $I \subset [m]$, $J \subset [n]$;
\REPEAT
\STATE Solve \eqref{eq:mred} and \eqref{eq:mred-dual} to obtain $x_J$ and $v_I$;
\STATE Choose $I_s \subset I_v := \{ i \in [m] \setminus I \, : \, A_{i,J}x_J < b_i \}$;
\STATE Choose $J_s \subset J_v := \{ j \in [n] \setminus J \, : \, A_{I,j}^T v_I > c_j \}$;
\STATE Set $I \leftarrow I \cup I_s$ and $J \leftarrow J \cup J_s$;
\UNTIL {$I_v=\emptyset$ and $J_v = \emptyset$;}
\STATE Set $x_{J^c} = 0$ and $v_{I^c} = 0$ and \textbf{terminate}.
\end{algorithmic}
\end{algorithm}


Many variants are possible within this framework. One could define the sets $I_s$ and $J_s$ to contain only the  smallest valid index, or the most-violated index. More commonly, $I_s$ and $J_s$ are chosen to have cardinality greater than $1$ where possible. In large-scale problems (analogous to partial pricing in the simplex method), not all checks in \eqref{eq:su8} are even performed. When $A$ is too large to store in memory, for example, we can stream columns of $A$ to calculate the quantities  $A_{I,j}^Tv_I - c_j$ until enough have been calculated to define $J_v$.


When Algorithm~\ref{alg:ccgen} is implemented with $I=[m]$ but $J$ a strict subset of $[n]$, it reduces to column generation. (In this case, $I_s$ and $I_v$ are null at every iteration.) Similarly, when $I$ is a strict subset of $[m]$ but $J=[n]$, Algorithm~\ref{alg:ccgen} reduces to constraint generation.

The success of constraint and column generation hinges on the ability to generate initial guesses for $I$ and $J$ that requires few additional iterations of Algorithm~\ref{alg:ccgen} to identify the solutions of \eqref{eq:mfull} and \eqref{eq:mfull-dual}.

\paragraph{The DantzigLP Framework.} Combining column and constraint generation LP techniques with methods for finding good initializations for the initial column and constraint sets $I$ and $J$, we develop DantzigLP, a general framework for solving large-scale versions of the the Dantzig Selector-type problems described in Section~\ref{section:introduction}. Initializations for $I$ and/or $J$ are obtained typically by solving the Lasso variant of a given problem.
This basic approach can be tailored to a large range of problems, and in many cases the problem structure admits fast algorithms for both initialization and column and constraint generation.

All of the Dantzig-type problems described in Section~\ref{section:introduction} (except for Basis Pursuit) have a tunable regularization parameter $\lambda$. Practitioners often wish to compute the estimator for a grid of $\lambda$ values specified a-priori. DantzigLP makes this process efficient by leveraging a simplex-based solver's warm start capabilities. Given the solution for one value of $\lambda$, DantzigLP can efficiently find the solution for its neighboring value of $\lambda$ (within the column and constraint generation framework). Repeating this process yields a path of solutions.



\subsection{The Dantzig Selector} \label{section:dantzig}

We show how the DantzigLP framework applies to $\DSLp$. We present an LP formulation for \DSL---the primal~\eqref{eq:candes_lp-1} and its corresponding dual~\eqref{eq:candes_lp-1-dual} are as follows: 

\begin{minipage}[c]{0.48\textwidth}
  \begin{equation}
    \label{eq:candes_lp-1}
    \begin{aligned}
    \underset{\beta^+, \beta^-, r}{\text{minimize}}~&~\sum_{i\in[p]} ( \beta^+_i + \beta^-_i ) \\
    ~ \sbt ~&~ - \M{1}\lambda \leq X^Tr  \leq \lambda \M{1} \\
     & r = y - X(\beta^+ - \beta^-) \\
    &  \beta^+, \beta^- \geq 0 \\
    \end{aligned}
  \end{equation}
\end{minipage}
\begin{minipage}[c]{0.01\textwidth}
~~~~
\end{minipage}
\begin{minipage}[c]{0.48\textwidth}
\begin{equation} \label{eq:candes_lp-1-dual}
\begin{aligned}
\maxi_{\nu^+,\nu^-,\alpha} ~&~ - \sum_{i\in[p]} \lambda (\nu^+_{i} + \nu^-_i) -  \alpha^T y \\
~ \sbt ~&~  - \M{1} \leq X^T\alpha \leq \M{1} \\
&   X(\nu^+ - \nu^-) + \alpha = 0 \\
& \nu^+, \nu^- \geq 0. \\
\end{aligned}
\end{equation}
\end{minipage}

The primal problem~\eqref{eq:candes_lp-1} has decision variables $\beta^+, \beta^{-} \in \mathbb{R}^{p}$ denoting the positive and negative parts of $\beta$, and
$r \in \mathbb{R}^{n}$ corresponding to the residual vector. The dual variables $\nu^+,\nu^- \in \mathbb{R}^{p}$ correspond to the inequality constraints $- \lambda \leq X^Tr$ and 
 $X^Tr\leq \lambda$ respectively, and $\alpha$ corresponds to the equality constraint $ r = y - X(\beta^+ - \beta^-)$. 
At optimality, the following complementarity conditions hold:
 $$ \nu^+ \circ ( X^T\alpha - \lambda \M{1}) = 0 ~~~~\text{and}~~~  \nu^- \circ (X'\alpha + \lambda \M{1}) = 0,$$
where, ``$\circ$" denotes componentwise multiplication.
Therefore, $X_{*,i}^Tr < \lambda \implies \nu^+_{i} = 0$ and $-X_{*,i}^Tr < \lambda \implies \nu^-_{i} = 0$.
Formulation~\eqref{eq:candes_lp-1} does not require computation and storage of the 
memory-intensive $p \times p$ matrix $X^TX$; this is avoided by introducing auxiliary variable $r$. Moreover, the related Lasso problem \eqref{eq:lasso} gives us reason to expect that this problem is a good candidate for both constraint and column generation. Optimality conditions for solution $\beta^L$ of \eqref{eq:lasso} can be written as follows:
\begin{equation}
    \label{eq:lasso-opt}
    r^L := y-X \beta^L, \quad
    X_{*,j}^T r^L \in \begin{cases}
    \{ \lambda \} & \;\; \mbox{if $\beta^L_j>0$} \\
    [-\lambda,\lambda] & \;\; \mbox{if $\beta^L_j =0$} \\
    \{-\lambda\} & \;\; \mbox{if $\beta^L_j<0$.}
    \end{cases}
\end{equation}
This suggests that, for the Lasso solution $\beta^L$ at least, the number of active constraints in \eqref{eq:candes_lp-1} is similar to the number of nonzero components in $\beta^L$, which is typically small. If the solution of the Dantzig selector has similar properties to the Lasso solution, then we would expect both the number of nonzero components in the solution of \eqref{eq:candes_lp-1} and the number of active constraints to be small relative to the dimensions of the problem.
(The papers \cite{james2009} and \cite{asif2010} demonstrate conditions under which the Dantzig and Lasso solution paths, traced as functions of the regularization parameter $\lambda$, are in fact identical.)



For a subset $J \subset [p]$ of columns of $X$ and $I \subset [p]$ of rows of $X$ (note that $I$ and $J$ need not be the same), we define
a reduced column and constraint version of \eqref{eq:candes_lp-1} as follows:
  \begin{equation}
    \label{eq:dantzig_lp}
    \tag{$\text{DS}(I, J)$}
    \begin{aligned}
     \mini_{\beta^+, \, \beta^-,\, r}~~ &~~
      \sum_{j \in J} \Big( \beta^+_j + \beta^-_j \Big) \\
\sbt~~& \; y - X_{*, J} (\beta^+_{J} - \beta^-_{J}) = r\\
&    (X_{*, I})^Tr  \leq \lambda  \M{1} \\
&    (X_{*, I})^Tr  \geq - \lambda \M{1} \\
&    \beta^+, \beta^-  \geq 0.
    \end{aligned}
  \end{equation}
Our constraint and column generation strategy for solving $\DSLp$ solves problems of this form at each iteration, initializing $I$ and $J$ from the Lasso solution, and alternately expanding $I$ and $J$ until a solution of $\DSL$ is identified. We initialize $J$ to be the subset of components $j \in [p]$ for which $\beta_j \neq 0$, while $I$ is the set of indices $i \in [p]$ such that $|X_{*,i}^Tr|=\lambda$. The strategy is specified as Algorithm~\ref{alg:ccgen.dslp}.


\begin{algorithm}
\caption{Constraint and Column Generation to solve $\DSLp$ \label{alg:ccgen.dslp}}
\begin{algorithmic}
\STATE Solve \eqref{eq:lasso} to obtain initial $I$ and $J$;
\LOOP
\STATE Calculate $I_v \leftarrow \{ i \in [p] \setminus I \, : \, |X_{*,i}^Tr|>\lambda \}$;
\IF{$I_v \neq \emptyset$}
\STATE Choose $\emptyset \neq I_s \subset I_v$; Set $I \leftarrow I \cup I_s$; Solve $\text{DS}(I,J)$;
\ELSE
\STATE Using $\alpha$ from the dual solution of $\text{DS}(I,J)$, calculate $J_v \leftarrow \{ j \in [p] \setminus J \, : \, |X_{*,j}^T \alpha| >1\}$;
\IF{$J_v \neq \emptyset$}
\STATE Choose $\emptyset \neq J_s \subset J_v$; Set $J \leftarrow J \cup J_s$; Solve $\text{DS}(I,J)$;
\ELSE
\STATE terminate.
\ENDIF
\ENDIF
\ENDLOOP
\end{algorithmic}
\end{algorithm}

Note that each time we solve $\text{DS}(I,J)$, we warm-start from the previous instance. The violation checks that define $I_v$ and $J_v$ can be replaced by relaxed versions involving a tolerance $\epsilon$. That is, we can define
\begin{equation}
I_v \leftarrow \{ i \in [p] \setminus I \, : \, |X_{*,i}^Tr|>\lambda +\epsilon\}, \quad
    J_v \leftarrow \{ j \in [p] \setminus J \, : \, |X_{*,j}^T \alpha| >1+\epsilon \}.
\end{equation}
 (We use $\epsilon=10^{-4}$ in our experiments.)

\paragraph{Computing a path of solutions.}  We can extend Algorithm~3 to solve $\DSL$ for a path of $\lambda$ values of the form $\Lambda = \{ \lambda_i, i \in [k] \}$ in decreasing order. We first obtain the Lasso solution for the smallest $\lambda$ value, which typically  corresponds to the densest solution in the Lasso path. (This strategy, which is the opposite of that used in solvers for Lasso --- see for example \cite{wright2009sparse} ---  reduces the overhead of continuously updating our LP model with new columns and constraints as we move across the $\lambda$-path.) The Lasso solution can be used to supply initial guesses of index sets  $I_0$ and $J_0$ for the first value $\lambda_1$. The final index sets $I_1$ and $J_1$ for $\lambda_1$ can then be used as initial guesses for $\lambda_2$, and so on. Optimal basis information (basis matrices and their factorizations) for each value of $\lambda$ can also be carried over to the next value.

\paragraph{Existing approaches.}
Many interesting approaches have been presented to solve the $\DSL$ problem.
\cite{becker2011} presents a general framework for by solving a regularized version of the problem with first order gradient based methods, but these  methods do not scale well.
\cite{lu2012,wang2012} use ADMM for $\DSLp$, which may lead to large feasibility violations. 
Homotopy methods were presented in
\cite{pang2017}  and were shown to outperforms existing algorithms for $\DSLp$.
Other homotopy based methods of appear in \cite{asif2009,brauer2018primal}. All the homotopy algorithms~\cite{asif2009,brauer2018primal,pang2017} compute the matrix
$X^TX$ at the outset; this memory intensive computation precludes the possibility of solving large scale instances $p\approx 10^6$ with $p \gg n$. Column and constraint generation has not been considered in these aforementioned papers.  

Computational results of our methods are presented in Section~\ref{sec:computation}.



\subsection{Basis Pursuit} \label{sec:basis-pursuit}
We study how our DantzigLP framework can be applied to \eqref{eq:basis_pursuit}, which admits the following LP representation. (Recall that we assume that the feasible set is nonempty.)
  \begin{equation}
    \label{eq:basis_pursuit_lp}
    \tag{\texttt{BP-Full}}
     \mini_{\beta^+, \beta^-}~~\sum_{j \in [p]} ( \beta^+_j + \beta^-_j ) ~~~~~ \sbt~~~~~y = X (\beta^+ - \beta^-), ~~ \beta^+ \geq 0, \;\; \beta^- \geq 0.
  \end{equation}  
Consider a subset of features $J \subset [p]$ and a restriction of \eqref{eq:basis_pursuit_lp} to the indices in $J$. We obtain the following  reduced primal~\eqref{eq:basis_pursuit_reduced} and  dual~\eqref{eq:basis_pursuit_dual}: 

\noindent
\begin{minipage}[c]{0.50\textwidth}
  \begin{equation}
    \label{eq:basis_pursuit_reduced}
    \tag{BPP(J)}
    \begin{aligned}
      & {\underset{\beta^+_J, \beta^-_J}{\text{minimize}}} & & \sum_{j \in J} (\beta^+_j + \beta^-_j) \\
      & \sbt
      & & y = X_{*, J} (\beta^+_J - \beta^-_J)  \\
      & & & \beta^+_J, \beta^-_J \geq 0
    \end{aligned}
  \end{equation}
\end{minipage}
\begin{minipage}[c]{0.02\textwidth}
~~
\end{minipage}
\begin{minipage}[c]{0.47\textwidth}
\begin{equation}
    \label{eq:basis_pursuit_dual}  \tag{BPD(J)}
    \begin{aligned}
      & {\underset{v}{\text{maximize}}} & & v^Ty  \\
      & \sbt
      & & X_{*, J}^Tv \leq \M{1}  \\
      & & & - X_{*, J}^Tv \leq \M{1}.  
   \end{aligned}
\end{equation}
\end{minipage}

For the column generation procedure, we need a subset $J$ (preferably of small size) for which~\eqref{eq:basis_pursuit_reduced} is feasible. Accordingly, we seek an approximation to the largest value of $\lambda$ for which the Lasso yields a a feasible solution for~\eqref{eq:basis_pursuit_reduced}. We find such a value by solving the Lasso for a sequence of decreasing values of $\lambda$, checking after each solution whether the resulting solution is feasible for~\eqref{eq:basis_pursuit_reduced}, and if so, defining $J$ to be the support obtained of this solution. 

If we cannot find a set $J$ for which~\eqref{eq:basis_pursuit_reduced} is feasible, we append the current $J$ with an additional $n-|J|$ columns to obtain a feasible 
solution\footnote{If the entries of $X$ are drawn from a continuous distribution, then $|J| = n$ will lead to a feasible solution for 
 \eqref{eq:basis_pursuit_reduced}. In all our experiments, the Lasso continuation approach did lead to a $J$ for which~\eqref{eq:basis_pursuit_reduced} was feasible. Note that the Lasso path often leads to solutions for which the number of nonzeros exceeds $n$.}.
 As before, the Lasso continuation approach is used just to obtain a good initialization for $J$ for our column generation framework, not to obtain a solution for BP.
 
 The approach is summarized in Algorithm~\ref{alg:ccgen.bp}; and computational results are presented in Section~\ref{sec:computation}.
 
\begin{algorithm}
\caption{Column Generation for Basis Pursuit~\eqref{eq:basis_pursuit_lp}\label{alg:ccgen.bp}}
\begin{algorithmic}
\STATE Solve a sequence of Lasso problems \eqref{eq:lasso} to obtain initial $J$;
\LOOP
\STATE Solve \eqref{eq:basis_pursuit_reduced}, with $v \in \mathbb{R}^n$ as the dual solution;
\STATE Calculate $J_v := \{ j \in [p] \setminus J \, : \, | X_{*,J}^Tv| > 1 \}$;
\IF{$J_v=\emptyset$} \STATE{terminate;}
\ELSE
\STATE Choose $\emptyset \neq J_s \subset J_v$ and set $J \leftarrow J \cup J_s$;
\ENDIF
\ENDLOOP
\end{algorithmic}
\end{algorithm}

\subsection{The Fused Dantzig Selector} \label{section:fused_dantzig}

\subsubsection{Signal estimation}\label{sec:signal-est-1}
We discuss how the DantzigLP framework can be extended to the Dantzig analog of \eqref{eq:fused_lasso} with $X=\mathbb{I}$ (the identity matrix), which is
\begin{equation}\label{eq-fused-lasso-id}
\frac12 \| y - \beta \|_{2}^2 + \lambda \| D^{(0)} \beta\|_{1},
\end{equation}
where $D^{(0)}$ is defined in \eqref{def.D0}.
To express this problem  in Lasso form, we define the $n \times n$ matrix
$D = [  e_1^T; D^{(0)}]$ (where $e_1=(1,0,0,\dotsc,0)^T$), which has full rank. Its inverse $H=D^{-1}$ is
\begin{equation}
    \label{eq:defn_H}
    H_{i, j} = \begin{cases}  
        1 &\text{if} \quad j = 1 \\
        i - j &\text{if} \quad i > j \\
        0 &\text{otherwise.}
    \end{cases}
\end{equation}
We can now rewrite \eqref{eq:fused_lasso} in terms of the variables $\alpha := D \beta$, and recover the solution $\beta$ of \eqref{eq-fused-lasso-id} by setting $\beta = H \alpha$. Defining $A := \{1\}$ and $B := \{2, \ldots, n\}$, we write \eqref{eq-fused-lasso-id} as follows:
\begin{equation}\label{fused-lasso-form-a-b}
\begin{aligned}
\mini_{\alpha_A, \alpha_B}~\frac{1}{2}\|y - H_A \alpha_A - H_B \alpha_B \|_2^2 + \lambda \| \alpha_B \|_1.
\end{aligned}
\end{equation}
This formulation differs slightly from the standard Lasso problem in that the  $\ell_1$ penalty term excludes $\alpha_A$.
The Dantzig analog of \eqref{fused-lasso-form-a-b} is
\begin{equation}
\begin{aligned}
\label{eq:dantzig_tf}
\mini_{\alpha_A, \alpha_B}~\|\alpha_B \|_1~~\sbt~~\| H_B^T(y - H_A \alpha_A - H_B \alpha_B) \|_\infty \leq \lambda,~H_A^T(y - H_A \alpha_A - H_B \alpha_B) = 0.
\end{aligned}
\end{equation}
(Note the constraint $H_A^T(y - H_A \alpha_A - H_B \alpha_B)=0$, which arises as an optimality condition for $\alpha_A$ in \eqref{fused-lasso-form-a-b}.)  Recalling that $H^{-1}=D$, and introducing auxiliary variables $\beta = H \alpha$, $r=y-\beta$, and $g=D^Tr$, we rewrite \eqref{eq:dantzig_tf} as follows:
 \begin{equation}
 \begin{aligned}
 \label{eq:fused_dantzig_compact}
       \mini_{r, \alpha^+, \alpha^-, \beta,\nabla} ~~~ & \sum_{i \in B} \big(\alpha_i^+ + \alpha_i^- \big) \\
        {\sbt}~~~~~ \beta &= D(\alpha^+ - \alpha^-) \\
      ~~r &= y - \beta \\
       ~~g &= D^T r \\
      ~~g_A &= 0 \\
      ~~g_B & \leq \lambda \M{1}  \\
      ~~g_B & \geq -\lambda \M{1} \\
      ~~~~\alpha^+, \alpha^-  & \geq 0.
 \end{aligned}
\end{equation}

We apply column and constraint generation to formulation~\eqref{eq:fused_dantzig_compact}: Column generation because of sparsity in $\alpha_B$, and constraint generation because few of the constraints $ g_{B} \in [-\lambda \M{1}, \lambda \M{1}]$ are expected to be active at optimality.
The formulation~\eqref{eq:fused_dantzig_compact} exploits the following key characteristics:
\begin{itemize}
    \item $H^{-1}=D$ is banded. Writing constraints in terms of $D$ rather than $H$ yields a constraint matrix with $O(n)$ nonzero entries. 
    (A direct LP representation for $\DSL$ with $n=p$ as in~\eqref{eq:candes_lp-1}
    would lead to a constraint matrix with $O(n^2)$ nonzeros.)
    \item Each component of $\alpha^+$ and $\alpha^-$ appears in exactly one equality constraint, reducing the cost of computing each reduced cost to $O(1)$ time ($\bar{c}_j = 1 - v_j$ for each $\alpha^+_j$, and $\bar{c}_j = 1 + v_j$ for each $\alpha^-_j$) --- much cheaper than the corresponding cost for a general \DSL~ problem, which requires the $O(n)$ operation $v^T X_j$.
    \item We can check each constraint violation $g_i \in [- \lambda,\lambda]$ in $O(1)$ time, compared to a general \DSL~problem which requires the $O(n)$ operation $X_{*,i}^Tr$ for each constraint.
\end{itemize}
To obtain a good initialization for \eqref{eq:fused_dantzig_compact}, we use the solution of \eqref{eq-fused-lasso-id}, which can be computed efficiently via dynamic programming~\cite{johnson2013} at $O(n)$ cost. 

\subsubsection{Beyond signal estimation: Regression}
We now consider problem \eqref{eq:fused_lasso} with general design matrix $X$.

 As in \eqref{fused-lasso-form-a-b}, we introduce a new variable $\alpha = H \beta$.
 We let $H_{A}$ be the submatrix of $H$ containing the columns indexed by $A$, and $P_{A}$ denote the projection operator onto the column space of $H_{A}$.
 With this notation in place, we rewrite~\eqref{eq:fused_lasso} in standard Lasso form
 with model matrix $\tilde{X} = (I - P_A) XH$ and response $\tilde{y} = (I - P_A) y$. 
The corresponding Dantzig Selector problem \eqref{eq:dantzig_tf_proj} is an instance of \DSL~problem with problem data $(\tilde{y}, \tilde{X})$. 
Since \eqref{eq:dantzig_tf_proj} lacks the structure as the signal estimation problem in Section~\ref{sec:signal-est-1} (where $X=\mathbb{I}$), we apply the DantzigLP procedure (Algorithm~3) to this problem, with a special initialization.
The initial point is obtained by solving \eqref{eq:fused_lasso} using a proximal gradient method~\cite{beck2009}. Each step of this method requires calculation of the proximal map defined by 
$$ \Theta(u_{k}) := \argmin_{\beta}~~\frac{L}{2} \left\| \beta - \left( u_{k} - \frac{1}{L} \nabla f( u_{k}) \right) \right\|_{2}^2 + \lambda \| D^{(0)}\beta \|_1, $$
where $L$ is the largest eigenvalue\footnote{This can be computed via the power method or by computing the largest singular value of $X$ with cost $O(\min\{n,p\}^2\max\{n,p\})$.} of $X^T X$ and $f(u) = 1/2\| y - Xu \|_{2}^2$. 
The  operator $\Theta(u_{k})$ is computed via dynamic programming~\cite{johnson2013}, which is highly efficient. If we set $u_{k}=\beta_{k}$ and $\beta_{k+1} = \Theta(\beta_{k})$ we get the usual (unaccelerated) proximal 
gradient algorithm. We use the accelerated variant which enjoys an improved convergence rate. It sets
$\tilde{u}_{k+1} = \Theta(u_{k})$ where $u_{k+1} = \tilde{u}_{k} + q_{k-1}(\tilde{u}_{k} - \tilde{u}_{k-1})/q_{k+1}$ and $q_{k+1} = (1 + \sqrt{1+4q_{k}^2})/2$. The sequence is initialized with
$u_{1} = \tilde{u}_{0}=0$ and $q_1=1$.

This proximal gradient approach to solving~\eqref{eq:fused_lasso} is much faster than a coordinate descent procedure \cite{friedman2010} applied to the Lasso reformulation of~\eqref{eq:fused_lasso} with problem data $(\tilde{y}, \tilde{X})$.


\section{Computational Results}\label{sec:computation}

This section presents computational results showing the performance of our proposed DantzigLP framework for the \DSL~problem (Section~\ref{sec:comp-DSL}), Basis Pursuit (Section~\ref{sec:comp-BP}),
and the Fused Dantzig Selector (Section~\ref{sec:comp-fused-DSL}).

\subsection{Computational experience with Dantzig Selector}\label{sec:comp-DSL}
We implement DantzigLP in Julia\footnote{Our Julia/JuMP implementation can be found at
\url{https://github.com/atzheng/DantzigLP}.}, using Gurobi's dual simplex solver\footnote{We use Gurobi version 7.5 in our experiments. All computations were performed on a Mac Pro desktop machine with specs: 2.7GHz 12-Core Intel Xeon E5. Unless otherwise specified the memory budget was 64GB of RAM.} as the LP solver and Lasso.jl (a Julia implementation of \texttt{glmnet}~\cite{friedman2010}) as the Lasso solver. At each iteration of column and constraint generation, we add up to 30 columns with the most negative reduced costs, and up to 50 of the most violated constraints. Unless stated otherwise, we solve problems to within a tolerance of $10^{-4}$ for both column and constraint generation violations.


\subsubsection{Experiments on synthetic datasets}\label{sec:DSL-synthetic}

\paragraph{Data Generation.}
Our first set of experiments were performed on synthetic data. The rows of $X$ are drawn from a multivariate Gaussian distribution $\text{MVN}(0, \Sigma)$ with mean zero and covariance $\Sigma$ where, 
$\Sigma_{ij}=\rho$ for all $i \neq j$ and $\Sigma_{ii}=1$. We then sparsify\footnote{Sparsification may destroy the correlation structure among columns of $X$.} $X$ by setting its entries to 0 independently with probability $\pi$; 
and finally normalize so that the columns of $X$ to have unit $\ell_2$-norm. 
To generate the true $\beta^0$, we choose a set $S \subset [p], |S| = {n}/{5}$ and set entries of $\beta^0_S$ as i.i.d. draws from a standard Gaussian distribution. The remaining components $\beta^0_{S^c}$ are set to zero. We then generate $y = X\beta^0 + e$ with $e_{i} \stackrel{\text{iid}}{\sim} N(0, \sigma^2)$; and $\sigma^2$ is chosen so as to achieve a signal-to-noise ratio (SNR) of 10. (Note that we define SNR as the ratio $\text{Var}{(X\beta)}/\sigma^2$.)



\begin{figure}
\includegraphics[width=\linewidth]{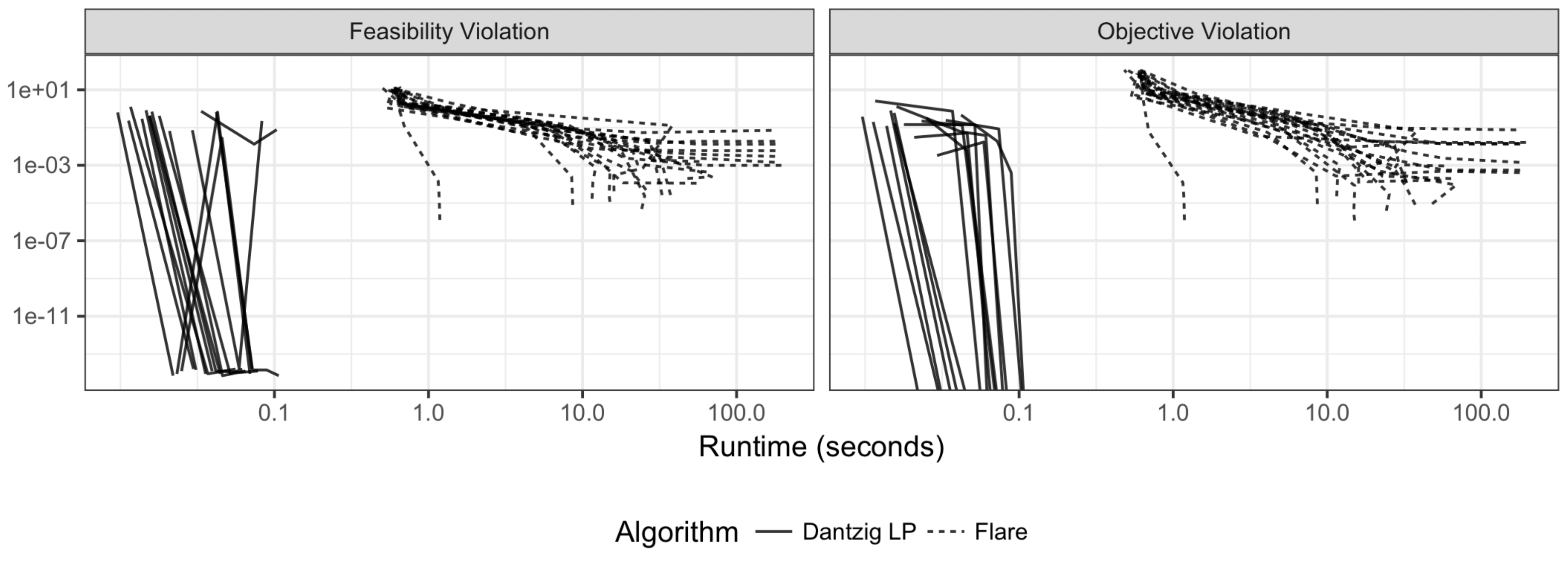}
    \caption{\small{Feasibility and objective violations as a function of runtime for DantzigLP (including time required for  Lasso initialization) and the ADMM implementation Flare, for the \DSL~ problem. 
    In all instances, DantzigLP reaches an optimal solution with zero feasibility violation within a few iterations. In comparison, Flare takes much longer to improve the objective and feasibility violations, often failing to converge even after hundreds of seconds.} }
\label{fig:admm_comparison}
\end{figure}

\paragraph{Comparison with ADMM.} 
A popular method for the \DSL~problem is based on ADMM~\cite{lu2012,wang2012,boyd2011}. In Figure~\ref{fig:admm_comparison}, we compare DantzigLP with \texttt{flare}~\cite{lu2012}, a publically available implementation of ADMM, plotting the violation in feasibility 
($\max \{ \| X^T(y - X\beta)\|_\infty - \lambda, 0 \}$) 
and the difference between the objective function from its optimal value ($| \| \beta\|_1 - \| \beta^*\|_1 |$) as a function of runtime.  (We use absolute value in the objective measure because infeasibility can result in a $\beta$ with smaller norm than the solution $\beta^*$.)  We find that ADMM is slow by both measures (an observation made also by \cite{pang2017}) and that DantzigLP is much faster.
 Each path in Figure~\ref{fig:admm_comparison} represents a single simulated problem instance with data generated by the means above, where $\rho=0$, $\pi=0$, $n = 200$, and $p = 1000$. We set $\lambda = \| X^T e^0 \|_\infty$ where, $e^0=(y-X \beta^0)$.
 Since ADMM has difficulty finding a solution with an absolute feasibility violation of less than 0.01 in many cases, we do not consider it further in the experiments below.

\paragraph{Comparison with PSM.}
The recently proposed parametric simplex-based solver PSM described in \cite{pang2017} can solve the $\DSLp$ problem and is a state-of-the-art solver for $\DSL$. 
In the next set of tests, we compare the following approaches.
\begin{enumerate}
    \item PSM, as implemented in the companion R package \texttt{fastclime}, for a path of 50 $\lambda$ values logarithmically spaced between $\lambda_{\text{min}} = 2 \|X^Te^0\|_\infty$ and $\lambda_{\text{max}} = \|X^Ty\|_\infty $.
    \item Gurobi (dual simplex method) applied to the full LP model~\eqref{eq:candes_lp-1} for $\lambda = \lambda_{\text{min}}$. We denote these results by ``Full LP (Single).''
    \item DantzigLP applied to \eqref{eq:candes_lp-1} for the same 50 $\lambda$ values as in the PSM tests. We denote these results by ``DantzigLP (Path).''
    \item DantzigLP applied to \eqref{eq:candes_lp-1} for $\lambda = \lambda_{\text{min}}$. We denote these results by ``DantzigLP (Single).''
\end{enumerate}
Note that PSM always computes a full path of solutions via homotopy, even if the solution is required at just one value of $\lambda$.
The times shown for our DantzigLP methods include the times required to compute a Lasso path,  
usually between 0.1--1 seconds\footnote{Solving the Lasso for $n=200,p=5000$ is the fastest with $n=1000,p=10^4$ being the slowest.}.

Table~\ref{tab:comparison} shows that when computing a path of 50 $\lambda$-values, PSM is usually outperformed by DantzigLP (Path). In computing a solution to \DSL~at a single $\lambda$, PSM is seen to be outperformed by solving the full LP model with Gurobi (denoted by ``Full LP''), with DantzigLP (Path) still faster.

\begin{table}[h!]
    \centering
\begin{tabular}{| llcccc|}
\hline
$n$ & $p$ & DantzigLP (Path) & DantzigLP (Single) & Full LP (Single) & PSM \\ 
\hline
200 & 5000  & $\phantom{0}0.81$ & $\phantom{0}0.12$ & $\phantom{0}5.1$ & $12.1$ \\
200 & 10000  & $\phantom{0}1.1$ & $\phantom{0}0.14$ & $10.3$ & $49.0$ \\
 \hline
 500& 5000  & $\phantom{0}4.5$ & $\phantom{0}1.1$ & $13.9$ & $16.0$ \\
 500& 10000  & $\phantom{0}5.9$ & $\phantom{0}1.3$ & $28.9$ & $86.5$ \\
 \hline
1000  & 5000  & $24.8$ & $\phantom{0}7.9$ & $31.9$ & $22.3$ \\
1000  & 10000  & $27.2$ & $\phantom{0}7.6$ & $65.1$ & $92.3$ \\
\hline 
\end{tabular}
    \caption{Runtime comparison for \DSL~ (synthetic instances) with $\rho=0$ and $\pi=0$. For each $n$ and $p$, we show the mean runtime (in seconds) of 20 problem instances. DantzigLP (with column and constraint generation) is usually faster than Gurobi's Full LP simplex solver (without column and constraint generation) and also faster than the state-of-the-art simplex-based homotopy solver PSM.
    }
    \label{tab:comparison}
\end{table}


We observe that PSM works well on instances with small $p$ (in the hundreds), but its performance deteriorates with increasing $p$ values. PSM computes the whole matrix $X^TX$, leading to large memory requirements by comparison with ``Full LP'' (which in our formulation does not require computation of $X^TX$) and also DantzigLP. Of all the methods, DantzigLP has the lowest memory requirements, as it generates new columns and constraints only as necessary.

\begin{table}[]
    \centering
   \scalebox{0.9}{\begin{tabular}{cc}
    \begin{tabular}{|ccccc|}
\hline
$\pi$ & DantzigLP & DantzigLP & Full LP & PSM\\
    &  (Path) & (Single) & (Single) & \\\hline
0 & 8.2 & 2.4 & 16.7 & 57.3\\
0.4 & 6.1 & 2.0 & 10.2 & 55.4\\
0.8 & 2.4 & 0.86 & 3.5 & 57.0\\
0.95 & 0.79 & 0.27 & 1.0 & 56.3\\
\hline
\end{tabular}&
    \begin{tabular}{|ccccc|}
\hline
    $\rho$ & DantzigLP & DantzigLP & Full LP & PSM\\
    &  (Path) & (Single) & (Single) & \\  \hline
0 & 8.2 & 2.4 & 16.7 & 57.3\\
0.4 & 3.2 & 0.57 & 12.7 & 15.4 \\
0.8 & 1.1 & 0.08 & 11.7 & 10.2\\
\hline
\end{tabular}
\end{tabular}}
    \caption{Runtimes (in seconds, averaged over 20 replications) for solving \DSL~ (synthetic instances) with $n = 500, p = 5000$ for varying sparsity and correlations in $X$. 
    [Left]  We vary sparsity in $X$, with $\rho=0$ and $\pi \in \{0, 0.4, 0.8, 0.95\}$ (larger $\pi$ values correspond to more zeroes in $X$). PSM has similar runtimes across all sparsity levels, whereas DantzigLP and Gurobi (Full LP) both see substantial reductions in runtime for sparse instances. [Right] We vary correlations in columns of $X$: $X$ is dense with $\pi=0$ and $\rho \in \{0, 0.4, 0.8\}$. Runtimes of all algorithms diminish with increasing correlations, with DantzigLP again fastest.}
    \label{tab:psm_sparsity}
\end{table}

\paragraph{Varying sparsity and correlation in $X$.} 
We next explore sensitivity of runtimes of the various approaches to sparsity in $X$ and correlations between columns of $X$, which are captured by the parameters  $\pi$ and $\rho$, respectively. Table~\ref{tab:psm_sparsity} (Left) shows that  the DantzigLP and Full LP approaches exploit sparsity  well (runtimes decrease with increasing sparsity), while
PSM does not benefit from sparsity, possibly becase the product $X^TX$ (which it computes) remains dense. 
For the case of dense $X$, Table~\ref{tab:psm_sparsity} (Right) shows that increasing correlations between the columns lead to improved runtimes for all algorithms. (The solutions $\beta$ tend to be sparser in these cases.) DantzigLP remains the clear winner.

\paragraph{Dependence on sparsity in solution $\beta$.} 
In the experiments above, we considered a sequence of $\lambda$-values in the range $[\lambda_{\min}, \lambda_{\max}]$ with $\lambda_{\min} = 2 \|X^Te^0\|_\infty$. It is well known that smaller 
values of  $\lambda$ correspond to denser solutions in $\beta$. To understand better the dependence of runtime of the various algorithms on sparsity of $\beta$, we tried the values
 $\lambda = \tau \|X^Te^0\|_\infty$, where $\tau \in \{0.1, 0.4, 0.7, 1\}$, showing the results in Table~\ref{tab:psm_lambda}. Runtimes for DantzigLP increase with density in $\beta$, as expected, mostly because the Lasso solution has a very different support from the solution of  $\DSL$, requiring DantzigLP to generate many more columns than it would for larger $\lambda$ values. For smaller values of $p$ (results not shown), DantzigLP can be even slower than the vanilla Gurobi implementation for the Full LP, due to the overhead of column generation, although memory consumption is still much smaller ($O(n^2)$ as opposed to $O(np)$). Computation time for PSM also increases as $\lambda$ decreases, though not as much as DantzigLP in relative terms. Still, DantzigLP performance remains superior for most interesting values of $\lambda$.

\begin{table}[h]
    \centering
  \begin{tabular}{|cccccc|}
\hline
$\tau$ & Avg. $L_0$ & DantzigLP & DantzigLP & Full LP & PSM\\
                   &                &  (Path) & (Single) & (Single) &\\  \hline
0.1 & 453. & 209. & 96.9 & 77.7 & 334.\\
0.4 & 320. & 79.2 & 30.7 & 74.5 & 368. \\
0.7 & 234. & 39.6 & 15.2 & 55.6 & 351.\\
1.0 & 182. & 13.6 & 4.7 & 24.1 & 125. \\
\hline
\end{tabular}
    \caption{Runtimes (in secs, averaged over 20 replications) for solving \DSL~ (synthetic instances) with $n = 500, p = 5000$,  $\rho=0$, $\pi=0$ for $\lambda_\tau = \tau \|X^Te^0\|_\infty$, where $\tau \in \{0.1, 0.4, 0.7, 1.0\}$.  DantzigLP (Path) solves for a path of 50 $\lambda$ values equally spaced between $\lambda_{\max}$ and $\lambda_\tau$; Full LP and DantzigLP (Single) solve only for $\lambda_\tau$; PSM solves for the whole path of values from $\lambda_{\max}$ to $\lambda_\tau$. The column labelled ``Avg. $L_0$'' shows average support size of the optimal solution to $\DSL$ at $\lambda_{\tau}$. As $\tau$ decreases, DantzigLP takes longer to solve the problem, although it still outperforms PSM.}  
    \label{tab:psm_lambda}
\end{table}

\begin{figure}[h!]
    \centering
    \includegraphics[scale=0.35]{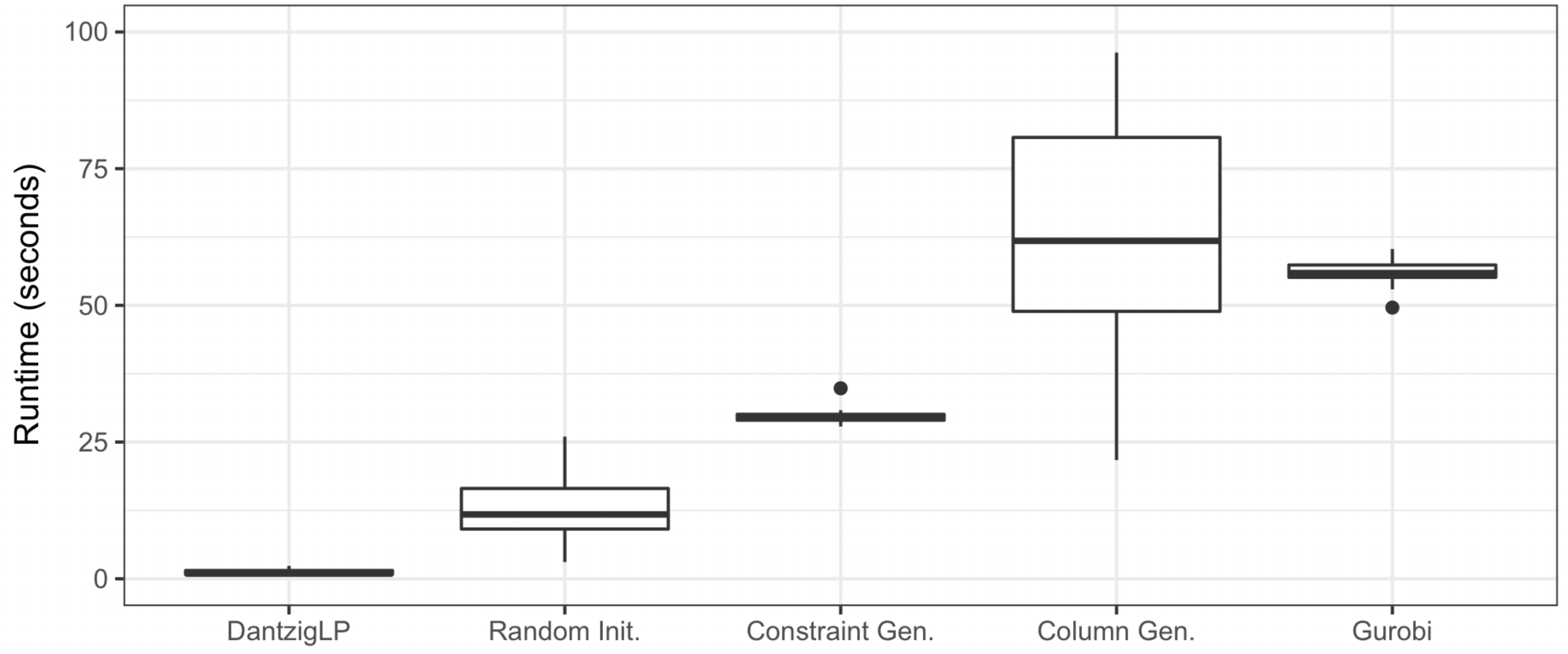}
    \caption{Comparison of 5 variants of DantzigLP ($n=1,000$, $p=10,000$), showing the importance of all  components of the approach:  Lasso initialization, column generation, constraint generation, and the simplex solver.
    \label{fig:colgen_vs_congen}}
\end{figure}

\paragraph{Importance of the components of DantzigLP.} 
The DantzigLP framework consists of several components: initialization provided by the Lasso, column generation and constraint generation, and the simplex LP solver. To understand the roles played by these components, we compare several variants in which some or other of them are omitted.  Figure~\ref{fig:colgen_vs_congen} compares the following five variants of DantzigLP.
\begin{enumerate}
    \item[(i)] {\em DantzigLP}: The complete framework described above.
    \item[(ii)] {\em Random Init.}: Rather than using a Lasso initialization, we initialize $I, J$ to a random subset of $[p]$ of size $\| \beta^0\|$.
    \item[(iii)] {\em Constraint Gen.}: Obtain $I$ from Lasso initialization, but $J = [p]$. That is, only constraint generation is enabled.
    \item[(iv)] {\em Column Gen.}: Here, $J$ is obtained from Lasso initialization, 
    with $I = [p]$. That is,  only column generation is enabled.
    \item[(v)] {\em Full LP}: The full LP model solved with Gurobi, without column or constraint generation. 
\end{enumerate}
The boxplots of Figure~\ref{fig:colgen_vs_congen} show the runtime distribution over 20 randomly generated \DSL~ instances with $n = 1,000$, $p=10,000$, $\rho=0$ and $\pi=0$, solved for $\lambda = 2\|X^Te^0\|_\infty$.
The figure highlights the importance of all elements of Dantzig LP.
We note that column generation alone provides no performance gains over the Full LP approach, due to the overhead of generating new columns (and restarting our simplex-based LP solvers). However, we see improvements when constraint generation is also introduced.

\subsubsection{Experiments on real datasets}
We demonstrate the performance of DantzigLP on real-datasets with $p \approx 10^6$ and $n\approx 10^3$. Due to memory constraints (our maximum was 64GB), only DantzigLP could solve these problems among our tested algorithms. We consider a path of 100 $\lambda$ values log-spaced in the interval 
$[\lambda_{\text{max}}, 10^{-3}\lambda_{\text{max}}].$ Results are displayed in Table~\ref{tbl:real_world}.

The ``AmazonLarge'' dataset from~\cite{hazimeh2018} has as its goal the prediction of helpfulness scores for product reviews based on data from Amazon's Grocery and Gourmet Food dataset. 
The ``Boston1M'' dataset consists of 104 covariates obtained from polynomial expansions of features in the Boston House Prices dataset (\cite{harrison1978}), augmented with 1000 random permutations of each column.



\begin{table}[h!]
    \centering
    \begin{tabular}{|l c cccc|}
\hline
Dataset & $\max ||\beta||_0$ & $n$ & $p$ & DantzigLP& Lasso \\
\hline
AmazonLarge & 178 & 2,500 & 174,755 & 248. & 70.7 \\
Boston1M & 56 & 200 & 1,000,103 & 3,702. & 44.3 \\
\hline
\end{tabular}
\caption{DantzigLP (and associated Lasso) runtimes in seconds on real datasets. We solve for a path of 100 $\lambda$ values. The column ``$\max \|\beta\|_0$" indicates the size of the support of the densest solution obtained. The DantzigLP runtime includes the runtime of the Lasso initialization step.}
\label{tbl:real_world}
\end{table}



\subsection{Results for Basis Pursuit}\label{sec:comp-BP}
We present some numerical results illustrating performance of the DantzigLP procedure for the basis pursuit problem \eqref{eq:basis_pursuit}. 
The matrix  $X$ is generated as in Section~\ref{sec:DSL-synthetic}, with $\rho=0$. We set $y = X \beta^0$, where $\beta^0$ is $0.2n$-sparse with nonzero elements chosen i.i.d. from $N(0,1)$. We compare DantzigLP against two other algorithms: the ADMM approach of \cite{boyd2011} as implemented in the ADMM R package, and solving the full LP model~\eqref{eq:basis_pursuit_lp}. In this example, we set a memory cap of 16GB of RAM for all experiments.

Table~\ref{tab:basis-pursuit} shows runtimes in seconds for a number of instances. We see that DantzigLP performs much better than competing algorithms in terms of runtime, particularly when $p\gg n$. This also comes with large savings in memory; the other algorithms were unable to solve problems of size $p \geq 10^5$ due to violation of the 16GB memory bound. DantzigLP can solve problems an order of magnitude larger than this. For instances with smaller $p$, the overhead of generating columns can cause DantzigLP to underperform the baselines in certain cases. The DantzigLP runtimes in Table~\ref{tab:basis-pursuit} include the runtime of the Lasso initialization step, which is the main performance bottleneck; the Lasso accounts for 80\% of the runtime for these instances.


While DantzigLP can obtain solutions of high accuracy, ADMM often has difficulty in doing so. The ADMM column in Table~\ref{tab:basis-pursuit} reports the minimum of time to converge to within $10^{-4}$ of the true objective and time to complete 10000 ADMM iterations (after which we terminate the algorithm). As the ``ADMM (\% Converged)'' column shows, for larger problem sizes none of the instances converge to that tolerance within the allotted iteration bound.


\begin{table}[]
    \centering
\begin{tabular}{|ccccccc|}
\hline
$n$ & $p$ & DantzigLP & Lasso & Full LP & ADMM & ADMM \\
      &  &          &          &     (Gurobi)              &           &    (\% Converged)\\
\hline
200 & $10^3$ & 1.9 & 0.38 & 1.3 & $>1.1$ & 90\\
200 & $10^5$ & 8.8 & 7.8 & NA & NA & NA\\
200 & $4 \times10^5$ & 31.3 & 28.8 & NA & NA & NA\\
\hline
500 & $10^3$  & 16.8 & 1.4 & 4.3 & 6.3 & 100\\
500 & $10^5$ & 25.3 & 19.1 & NA & NA & NA\\
500 & $4 \times10^5$ & 72.7 & 62.7 & NA & NA & NA\\
\hline
1000 & $10^3$  & 27.1 & 3.0 & 10.9 & 26.5 & 100\\
1000 & $10^5$ & 88.7 & 42.1 & NA & NA & NA\\
1000 &$4 \times 10^5$ & 209. & 116. & NA & NA & NA\\\hline
\end{tabular}
    \caption{Mean runtimes in seconds for simulated basis pursuit instances (20 instances per setting of $n$ and $p$), comparing DantzigLP, the Lasso initialization step, the Full LP implementation in Gurobi, and ADMM. The ``NA" values indicate cases where the algorithm was terminated due to memory constraints, with a maximum memory allocation of 16GB. ``ADMM (\% Converged)'' indicates the percentage of instances in which ADMM was able to converge. The ``ADMM'' column reports the minimum of time to termination and time to convergence, with a ``$>$" symbol indicating cases where fewer than 100\% of instances converged.}
    \label{tab:basis-pursuit}
\end{table}

\subsection{Computational experience with Fused Dantzig Selector}\label{sec:comp-fused-DSL}

\paragraph{Signal estimation.}
We first consider the Fused Dantzig Selector with $X=\mathbb{I}$, that is, the signal estimation case of
Section~\ref{sec:signal-est-1}. We generate a piecewise constant signal, with discontinuities / knots chosen at random from $[n]$. At each knot, the jump is chosen from $N(0,1)$. We add noise with SNR=10 to the signal. 
We solve \eqref{eq:fused_dantzig_compact} at a single value $\lambda = \| H_B^T(y - H_A \alpha^0_A - H_B \alpha^0_B) \|_\infty$, where $(\alpha^0_A,\alpha_B^0)$ corresponds to the true signal.

In Table~\ref{tab:fused-dantzig}, we compare our DantzigLP framework\footnote{We solve each instance to within a tolerance of $\epsilon=10^{-4}$ and add up to 40 columns (constraints) per iteration of column (constraint) generation.} to directly solving the full version \eqref{eq:fused_dantzig_compact}. Gurobi's dual simplex solver is used in both cases.
The formulation \eqref{eq:fused_dantzig_compact} allows solution of problems several orders of magnitude larger than the Dantzig Selector problem~\eqref{eq:candes_lp-1}, when  column/constraint generation is not used.
The DantzigLP framework improves modestly on this enhancement,  by factors of 2-3 for problems with large $n$ and a small number of knots. 
The runtime for the full LP formulation is  insensitive to the number of knots, while the runtime of DantzigLP increases with the number of knots with a large number of knots. This is not surprising, as more knots implies a denser solution.
Solving the Fused Lasso (required for initialization) takes less than one second across all instances.

\begin{table}[]
    \centering
\begin{tabular}{cc}
    \begin{tabular}{|cccc|}
\hline
$n$ & \# knots & DantzigLP & Full LP\\
\hline
$5\times10^4$ & 100 & 3.5 & 3.3\\
      & 200 & 3.5 & 3.2\\
      & 1,000 & 4.2 & 3.1\\
\hline
$10^5$  & 100 & 5.7 & 7.9\\
      & 200 & 6.2 & 8.0 \\
      & 1,000 & 8.2 & 7.9\\
      \hline
\end{tabular}&
    \begin{tabular}{|cccc|}
\hline
$n$ & \# knots & DantzigLP & Full LP\\
\hline
$2\times10^5$& 100 & 10.8 & 24.0\\
      & 200 & 11.9 & 24.1\\
      & 1,000 & 16.4 & 23.6\\
\hline
$5\times10^5$ & 100 & 48.7 & 120.3\\
      & 200 & 54.1 & 121.4\\
      & 1,000 & 80.9 & 120.6\\
      \hline
\end{tabular}
\end{tabular}
    \caption{[Signal Estimation] Runtimes (seconds) for the Fused Dantzig Selector, comparing DantzigLP with solution of the Full LP formulation \eqref{eq:fused_dantzig_compact} using Gurobi. Because of memory requirements, it would not be possible directly solve~\eqref{eq:dantzig_tf} without using our proposed reformulation \eqref{eq:fused_dantzig_compact} for the instances considered here.}
    \label{tab:fused-dantzig}
\end{table}

\begin{table}[]
    \centering
    \begin{tabular}{|c c c c|}
\hline
$n$ & $p$ & DantzigLP & Gurobi\\
\hline
500 & 5,000 & 33.3 & 82.8\\
500 & 10,000 & 117.3 & 341.7 \\
\hline
1,000 & 5,000 & 67.0 & 221.3\\
1,000 & 10,000 & 207.7 & 925.2\\
\hline
\end{tabular}

    \caption{[Regression] Runtimes (in seconds, averaged over 20 replications) for the fused Dantzig Selector with $X \neq I$. We compare our DantzigLP framework with Full LP (Gurobi).}
    \label{tab:fused_regression}
\end{table}

\paragraph{Regression.}
We illustrate the performance of our DantzigLP framework for \eqref{eq:dantzig_tf_proj} for a general model matrix $X \neq \mathbb{I}$.
Here, entries of $X$ are drawn from a standard Gaussian ensemble: i.e., $X_{ij} \stackrel{\text{iid}}{\sim} N(0,1)$.
The underlying $\beta^0$ is piecewise constant and is drawn as before with 20 knots, with jumps chosen i.i.d. from $N(0,1)$.
 We generate $y=X\beta^0 + e$, with $e_{i}  \stackrel{\text{iid}}{\sim} N(0,\sigma^2)$ and $\sigma$ chosen to produce an SNR of 10.

Table~\ref{tab:fused_regression} shows that the Fused Dantzig Selector modestly outperforms solving the full LP with Gurobi. The runtime improvements are not as pronounced as  in the case of \DSL~ because the initial solution obtained from the Lasso problem \eqref{eq:fused_lasso} is not as accurate as for \DSL. Thus, additional columns/constraints need to be generated to achieve optimality, leading to increased runtimes.

\section{Acknowledgements}
Rahul Mazumder acknowledges research support from the Office Naval Research ONR-N000141512342, ONR-N000141812298 (Young Investigator Award) and the National Science Foundation (NSF-IIS-1718258).

\bibliographystyle{plainnat_my_nourl}
\bibliography{dantzig_lp.bib}

\end{document}